\documentclass[preprint,superscriptaddress,showpacs,amsmath,amssymb,aps,pra]{revtex4-1}

\usepackage{amsthm}
\usepackage{amsmath}
\usepackage{amssymb}
\usepackage{braket}
\usepackage{bbold}
\usepackage{graphicx}
\usepackage{subfigure}
\usepackage{hyperref}
\usepackage{appendix}
\usepackage{tikz}
\usepgflibrary{shapes.geometric}
\usetikzlibrary{arrows}

\theoremstyle{definition}

\hypersetup{colorlinks=true, citecolor=blue, urlcolor=blue, linkcolor=blue}

\begin{document}
\title{Estimating Parameterized Entanglement Measure}
\author{Zhi-Wei Wei}
\email{weizhw@cnu.edu.cn (corresponding author)}
\affiliation{School of Mathematical Science,Capital Normal University,100048,Beijing,China}
\author{Ming-Xing Luo}
\email{mxluo@home.swjtu.edu.cn}
\affiliation{The School of Information Science and Technology, Southwest Jiaotong University, Chengdu 610031, China}
\author{Shao-Ming Fei}
\email{feishm@cnu.edu.cn (corresponding author)}
\affiliation{School of Mathematical Science,Capital Normal University,100048,Beijing,China}
\affiliation{Max-Planck-Institute for Mathematics in the Sciences, 04103 Leipzig, Germany}

\bigskip

\begin{abstract}
The parameterized entanglement monotone, the $q$-concurrence, is also a reasonable parameterized entanglement measure. By exploring the properties of the $q$-concurrence with respect to the positive partial transposition and realignment of density matrices, we present tight lower bounds of the $q$-concurrence for arbitrary $q\geqslant 2$. Detailed examples are given to show that the bounds are better than the previous ones.
\end{abstract}

\maketitle

\section{introduction}\label{introduction}
Entanglement is one of the most remarkable phenomena in quantum mechanics \cite{schrodinger1935gegenwartige,PhysRev.47.777}. In recent years, great efforts have been made toward the understanding of the role played by the entanglement in quantum information theory \cite{nielsen2002quantum}. It has been the most important resources in quantum information processing and communication such as quantum dense coding \cite{PhysRevLett.69.2881}, quantum metrology \cite{PhysRevA.46.R6797,PhysRevLett.96.010401}, quantum teleportation \cite{PhysRevLett.70.1895,PhysRevA.61.062306}, quantum secret sharing \cite{PhysRevA.59.1829} and quantum cryptography \cite{RevModPhys.74.145}. These applications have strongly motivated the study on detection and quantification of entanglement in an operational way.

Detecting the entanglement of generic mixed states is still a hard problem. The positive partial transpose $\left(\mathrm{PPT}\right)$ criterion \cite{PhysRevLett.77.1413} says that
for any separable bipartite state $\rho_{AB}$, the partial transposed matrix  $\rho^{\Gamma}\geqslant 0$ ($\Gamma$ represents the partial transposition with respect to the subsystem $B$ in the following) is semi-positive. The PPT criterion is a necessary and sufficient condition of separability only for pure states and $2\otimes 2$ and $2\otimes 3$ dimensional mixed states \cite{PhysRevLett.77.1413,HORODECKI19961}. While the realignment criterion \cite{PhysRevA.59.4206,rudolph2004computable,rudolph2005further,chen2003matrix} says that the realigned matrix $\mathcal{R}\left(\rho\right)$ of any separable $\rho_{AB}$ satisfies $\left\|\mathcal{R}\left(\rho\right)\right\|_1\leqslant 1$, where $\|X\|_1$ denotes the trace norm defined by $\|X\|_1=\mathrm{Tr}\sqrt{XX^{\dagger}}$.
These separability criteria have been widely used for entanglement detection both theoretically and experimentally in quantum information processing \cite{RevModPhys.81.865}.

The quantification of quantum entanglement for a given quantum state is also a difficult undertaking due to the intricate interplay between classical and quantum correlations \cite{PhysRevLett.78.2275,PhysRevA.57.1619}. It has been proposed that a reasonable measure of entanglement should fulfill certain conditions \cite{PhysRevLett.78.2275,PhysRevA.57.1619,PhysRevA.59.141}. Some interesting entanglement measures have been provided for bipartite systems such as concurrence \cite{PhysRevLett.78.5022,PhysRevA.64.042315,PhysRevLett.95.040504}, entanglement of formation \cite{PhysRevA.54.3824,PhysRevLett.80.2245,horodecki2001entanglement}, negativity \cite{PhysRevA.65.032314}, robustness of entanglement \cite{PhysRevA.65.052327} and R\'enyi-$\alpha$ entropy of entanglement \cite{PhysRevA.93.022324}. There are also some entanglement monotones \cite{gventangledmontone} such as the convex-roof extension of negativity \cite{PhysRevA.68.062304}, Tsallis-$q$ entropy of entanglement \cite{PhysRevA.81.062328}, as well as the entanglement monotones induced by fidelity \cite{guo2020entanglement}.

Nevertheless, most proposed entanglement measures or monotones involve extremizations which are difficult to handle analytically. Usually analytical results are only available for two-qubit states \cite{PhysRevLett.80.2245} or some special higher-dimensional mixed states \cite{PhysRevLett.85.2625,PhysRevA.64.062307,PhysRevA.67.012307} for certain special measures \cite{PhysRevA.68.062304,buchholz2016evaluating}. Therefore, efforts have been made towards the estimation of entanglement measures for general mixed states \cite{PhysRevLett.95.210501}. The analytical lower bound for the concurrence has been derived in \cite{PhysRevLett.95.040504} based on the PPT and realignment criteria.
In \cite{Li_Guo_2009} the authors sharpened this bound by relating the concurrence to the local uncertainty relations and the correlation matrix criterion.
By using the PPT and realignment criteria, a lower bound for the genuine tripartite entanglement concurrence was obtained in \cite{li2017detection}.

In particular, the concurrence plays a vital role in entanglement distributions such as entanglement swapping and remote preparation of bipartite entangled states \cite{PRL93260501}.
For any pure state $\ket{\Psi}_{AB}$ the concurrence is given by $C(\ket{\Psi}_{AB})= \sqrt{2(1-\mathrm{Tr}\rho_{A}^2)}$ with $\rho_A=\mathrm{Tr}_B|\Psi\rangle\langle\Psi|$ \cite{PhysRevLett.78.5022}. In fact, the concurrence is related to the specific Tsallis-$q$ entropy $T_q(\rho_{A})$ with $q=2$, $C(\ket{\Psi}_{AB})= \sqrt{2T_2(\rho_{A})}$ \cite{tsallis1988possible,LANDSBERG1998211}. Noteworthily, the Tsallis-$q$ entropy provides a generalization of the traditional Boltzmann-Gibbs statistical mechanics and enables one to find a consistent treatment of dynamics in many nonextensive physical systems such as with long-range interactions, long-time memories and multifractal structures \cite{abe2001nonextensive}. The Tsallis-$q$ entropy also provides many intriguing applications in the realms of quantum information theory \cite{ABE2001157,PRA63042104,PRA66042306,PRA72022322}. Therefore, it is of great significance to provide an entanglement measure from the Tsallis entropy with $q>2$. Recently, the authors in \cite{PhysRevA.103.052423} has presented such parameterized entanglement monotone for $q>2$.

In this paper we present analytical tighter lower bounds for the parameterized entanglement monotone $q$-concurrence given in \cite{PhysRevA.103.052423}, which is also a well-defined entanglement measure. The rest of this paper is organized as follows. In Sec. \ref{boundq}, we recall some necessary conditions for bipartite entanglement measures, and the concept of $q$-concurrence. We derive tighter lower bounds of the $q$-concurrence for general mixed states, and consider some detailed examples in Sec. \ref{qisotropics}. We make a conclusion in Sec. \ref{conclusion}.

\section{bound on $q$-concurrence}\label{boundq}
Let $\mathcal{D}$ denote the set of bipartite states in finite dimensional Hilbert space $\mathcal{H}_A\otimes\mathcal{H}_B$ associated with subsystems $A$ and $B$. A well-defined quantum entanglement measure $E$ must satisfy certain conditions \cite{PhysRevLett.78.2275,PhysRevA.57.1619,PhysRevA.59.141} as follows:
\begin{itemize}
\item[(i)] $E\left(\rho\right)\geqslant0$ for any state $\rho\in\mathcal{D}$, where the equality holds only for separable states.
\item[(ii)] $E$ is invariant under local unitary transformations, $E\left(\rho\right)=E\left(U_A\otimes U_B\rho U_A^{\dagger}\otimes U_B^{\dagger}\right)$ for any local unitaries $U_A$ and $U_B$.
\item[(iii)] $E$ does not increase on average under stochastic LOCC,
\begin{equation}\label{avm}
E\left(\rho\right)\geqslant\sum_ip_iE\left(\rho_i\right)
\end{equation}
for any $\rho\in\mathcal{D}$, where $p_i=\mathrm{Tr}A_i\rho A_i^{\dagger}$ is the probability of obtaining outcome $i$, and $\rho_i=A_i\rho A_i^{\dagger}/p_i$ with $A_i$ the Kraus operators with respect to the stochastic LOCC such that $\sum_iA_i^{\dagger}A_i=I$.
\item[(iv)] $E$ is convex,
\begin{align}\label{con}
E\left(\sum_ip_i\rho_i\right)\leqslant\sum_ip_iE\left(\rho_i\right).
\end{align}
\item[(v)] $E$ cannot increase under LOCC, $E\left(\rho\right)\geqslant E\left(\Lambda\left(\rho\right)\right)$ for any LOCC map $\Lambda$.
\end{itemize}

The condition (ii) can be removed if the condition (v) holds.
$E$ is said to be an entanglement monotone \cite{gventangledmontone} if the first four conditions hold. In \cite{horodecki2005simplifying}, it has been shown that a convex function $E$ satisfies condition (v) if and only if it satisfies condition (ii) and
\begin{equation}\label{preliminary1}
E\left(\sum_ip_i|i\rangle\langle i|_M\otimes\rho_i\right)=\sum_ip_iE\left(\rho_i\right),
\end{equation}
where $M=A', B'$ is a flag system and $\{\ket{i}\}$ are the local orthogonal basic vectors. In addition, (\ref{preliminary1}) is just the flag additivity which is equivalent to the average monotonicity and the convexity, i.e., the conditions (iii) and (iv) \cite{PhysRevA99042322}. In this case, any entanglement monotone defined in \cite{gventangledmontone} is an entanglement measure.

Any pure state $\ket{\psi}\in\mathcal{H}_A\otimes\mathcal{H}_B$ can be expressed in the Schmidt form under suitable local bases $\{\ket{i_A}\}$ and $\{\ket{i_B}\}$ of $\mathcal{H}_A$ and $\mathcal{H}_B$, respectively,
\begin{equation}
\label{mrt1}
\ket{\psi}_{AB}=\sum_{i=1}^d\sqrt{\lambda_i}\ket{i_A\,i_B},
\end{equation}
where $\lambda_i$'s are the squared Schmidt coefficients with $\sum_{i=1}^d\lambda_i=1$, and $d=\min\set{d_A,d_B}$ with $d_A$ and $d_B$ the dimensions of $\mathcal{H}_A$ and $\mathcal{H}_B$, respectively \cite{nielsen2002quantum}.
The parameterized entanglement monotone $q$-concurrence $C_q\left(\ket{\psi}_{AB}\right)$ of a state $\ket{\psi}_{AB}$ \cite{PhysRevA.103.052423} is defined by
\begin{equation}\label{textp3}
C_q\left(\ket{\psi}_{AB}\right)=1-\mathrm{Tr}\rho_A^q
\end{equation}
for any $q\geqslant 2$, where $\rho_A=\mathrm{Tr}_B\left(|\psi\rangle\langle\psi|\right)$ is the reduced density operator of the subsystem $A$. It is direct to verify that $C_q\left(\ket{\psi}_{AB}\right)=1-\sum_{i=1}^d\lambda_i^q$ and $0\leqslant C_q\left(\ket{\psi}\right)\leqslant1-d^{1-q}$, where the left equality holds if $\ket{\psi}$ is a product state, and the right equality holds for the maximally entangled state $\ket{\psi}=1/\sqrt{d}\sum_i\ket{ii}$. The $q$-concurrence for general mixed states $\rho\in\mathcal{D}$ is given by convex-roof extension,
\begin{equation}\label{text002}
C_q\left(\rho\right)=\min_{\left\{p_i,\ket{\psi_i}\right\}}
\sum_ip_iC_q\left(\ket{\psi_i}\right),
\end{equation}
where the infimum is taken over all possible pure state decompositions of $\rho=\sum_ip_i|\psi_i\rangle\langle\psi_i|$, with $\sum_ip_i=1$ and $p_i\geqslant 0$.

It has been proved in \cite{PhysRevA.103.052423} that the $q$-concurrence $C_q\left(\rho\right)$ is already an entanglement monotone. Therefore, it is also an entanglement measure.

It is hard to derive an analytic formula of the entanglement measure $q$-concurrence for general mixed states. In the following we estimate the $q$-concurrence by deriving its lower bounds based on the PPT and realignment criteria. For a given bipartite state $\rho=\sum_{ijkl}\rho_{ij,kl}|ij\rangle\langle kl|$ in computational bases, the partial transposed matrix $\rho^{\Gamma}$ with respect to the subsystem $B$ is given by $\rho^{\Gamma}=\sum_{ijkl}\rho_{ij,kl}|il\rangle\langle kj|$, and the realigned matrix is given by $\mathcal{R}\left(\rho\right)=\sum_{ijkl}\rho_{ij,kl}|ik\rangle\langle jl|$. For a given pure state $\ket{\psi}$ defined in (\ref{mrt1}), it is straightforward to show that \cite{PhysRevLett.95.040504},
\begin{equation}\label{text1}
1\leqslant \left\|\rho^{\Gamma}\right\|_1=\left\|\mathcal{R}\left(\rho\right)\right\|_1
=\left(\sum_{i=1}^d\sqrt{\lambda_i}\right)^2\leqslant d,
\end{equation}
where $\rho=|\psi\rangle\langle\psi|$.
In particular, for $q=2$ the $q$-concurrence becomes $C_2\left(\ket{\psi}\right)=1-\sum_{i=1}^d\lambda_i^2=2\sum_{i<j}\lambda_i\lambda_j$. One has then \cite{PhysRevLett.95.040504},
\begin{equation}\label{b2p}
C_2\left(\ket{\psi}\right)\geqslant\frac{1}{d\left(d-1\right)}
\left(\left\|\rho^{\Gamma}\right\|_1-1\right)^2
\end{equation}
for any pure state $\ket{\psi}$ on $\mathcal{H}_A\otimes\mathcal{H}_B$.

Before deriving a tight lower bound of $q$-concurrence for $q\geqslant2$, we first show the following conclusion. For a given pure state $\ket{\psi}$ defined in (\ref{mrt1}), let us analyze the monotonicity of the function $f\left(q\right)$ given by,
\begin{align}\label{bound1}
f\left(q\right)=\frac{1-\sum_{i=1}^d\lambda_i^q}{1-d^{1-q}}
\end{align}
for any $q\geqslant2$. Set
\begin{align}\label{bound2}
G_{dq}=\sum_i\lambda_i^q\ln\lambda_i\left(d^{1-q}-1\right)
-\left(1-\sum_i\lambda_i^q\right)d^{1-q}\ln d.
\end{align}
We have
\begin{equation}\label{text006}
\frac{\partial f}{\partial q}=\frac{G_{dq}}{\left(1-d^{1-q}\right)^2}.
\end{equation}
We see that the sign of the first derivative of $f\left(q\right)$ with respect to $q$ depends on the sign of the function $G_{dq}$, with constraints $\sum_{i=1}^d\lambda_i=1$ and  $\lambda_i>0$ for $i=1,...,d$. Consider the minimum of $G_{dq}$ by using Lagrange multipliers \cite{PhysRevA.67.012307} subject to the constraints $\sum_i\lambda_i=1$ and $\lambda_i>0$. There is only one stable point under the constraints, $\lambda_i=1/d$, $i=1,\cdots,d$, for which we have $G_{dq}=0$.

The second derivative of $f\left(q\right)$ with respect to $q$ at the stable point is given by
\begin{align}\label{bound3}
\frac{\partial^2G_{dq}}{\partial\lambda_i^2}\Big|_{\lambda_i
=\frac{1}{d}}=d^{2-q}\left[q\left(q-1\right)\ln d-\left(2q-1\right)\left(1-d^{1-q}\right)\right].
\end{align}
Therefore, we get
\begin{align}\label{bound4}
\frac{\partial^2G_{dq}}{\partial\lambda_i^2}\Big|_{\lambda_i=\frac{1}{d}}\geqslant0
\end{align}
for $q\geqslant s\equiv 2.4721$, $d=2$ and $q\geqslant2$, $d\geqslant3$. In these cases, the minimum extreme point is the minimum point and $\partial f/\partial q\geqslant 0$.
From the above analysis, it is straightforward to have the following conclusion.

$\mathit{Corollary\,1}$. If $q\geqslant h$, then  $C_q\left(\rho\right)\geqslant\frac{1-d^{1-q}}{1-d^{1-h}}C_h\left(\rho\right)$ for either $h\geqslant s$, $d=2$ or $h\geqslant2$, $d\geqslant3$.

We now derive the main result of this paper.

$\mathit{Theorem\,1}$.
For a general bipartite state $\rho\in\mathcal{D}$, we have
\begin{equation}
\label{text101}
C_q\left(\rho\right)\geqslant\frac{1-d^{1-q}}{\left(d-1\right)^2}
\left[\max\left(\left\|\rho^{\Gamma}\right\|_1,\left\|\mathcal{R}
\left(\rho\right)\right\|_1\right)-1\right]^2
\end{equation}
for either $q\geqslant2$ with $d\geqslant3$ or $q\geqslant3$ with $d=2$, and
\begin{align}\label{corbound4}
C_q\left(\rho\right)>\frac{1-2^{1-q}}{2-2^{2-s}}\left[\max
\left(\left\|\rho^{\Gamma}\right\|_1,\left\|\mathcal{R}
\left(\rho\right)\right\|_1\right)-1\right]^2
\end{align}
for $s\leqslant q<3$ with $d=2$.

$\mathit{Proof}$. For a given pure state (\ref{mrt1}), from (\ref{bound4}) and (\ref{text006}) we can obtain that $f(q)$ in (\ref{bound1}) is an increasing function with respect to $q$ in the cases of $q\geqslant2$ with $d\geqslant3$, and $q\geqslant s$ with $d=2$.

In the first case of $d\geqslant3$, we have
\begin{align}\label{theorem4}
C_q\left(\ket{\psi}\right)&\geqslant\frac{1-d^{1-q}}{1-d^{-1}}C_2
\left(\ket{\psi}\right)\nonumber\\
&\geqslant\frac{1-d^{1-q}}{\left(d-1\right)^2}\left(\|\sigma^{\Gamma}\|_1-1\right)^2
\end{align}
for $q\geqslant2$, where $\sigma=|\psi\rangle\langle\psi|$, the first inequality is due to the monotone increasing of $f\left(q\right)$, the second inequality is due to (\ref{b2p}).

For case of $d=2$, similar to (\ref{theorem4}) we can obtain when $q\geqslant3$,
\begin{align}\label{theor5}
C_q\left(\ket{\psi}\right)&\geqslant\frac{1-2^{1-q}}{1-2^{-2}}C_3
\left(\ket{\psi}\right)\nonumber\\
&=\left(1-2^{1-q}\right)2C_2\left(\ket{\psi}\right)\nonumber\\
&\geqslant\left(1-2^{1-q}\right)\left(\|\sigma^{\Gamma}\|_1-1\right)^2,
\end{align}
where the equality is due to $f\left(2\right)=f\left(3\right)$. When $s\leqslant q<3$, we have
\begin{align}\label{case1}
C_q\left(\ket{\psi}\right)&\geqslant\frac{1-2^{1-q}}{1-2^{1-s}}C_s\left(\ket{\psi}\right)\nonumber\\
&>\frac{1-2^{1-q}}{1-2^{1-s}}C_2\left(\ket{\psi}\right)\nonumber\\
&>\frac{1-2^{1-q}}{2-2^{2-s}}\left(\left\|\sigma^{\Gamma}\right\|_1-1\right)^2,
\end{align}
where the second inequality is from the monotonic increasing of $C_q\left(\ket{\psi}\right)$ with respect to $q$.

Let $\rho=\sum_{i}p_i|\psi_i\rangle\langle\psi_i|$ be the optimal pure state decomposition of $C_q\left(\rho\right)$ for a given mixed state $\rho\in\mathcal{D}$. For the cases of $q\geqslant2$ with $d\geqslant3$ and $q\geqslant3$ with $d=2$, we have
\begin{align}\label{text9}
C_q\left(\rho\right)&=\sum_ip_iC_q\left(\ket{\psi_i}\right)\nonumber\\
&\geqslant \frac{1-d^{1-q}}{\left(d-1\right)^2}\sum_ip_i\left(\left\|\rho_i^{\Gamma}\right\|_1-1\right)^2\nonumber\\
&\geqslant \frac{1-d^{1-q}}{\left(d-1\right)^2}\left(\sum_ip_i\left\|\rho_i^{\Gamma}\right\|_1-1\right)^2\nonumber\\
&\geqslant  \frac{1-d^{1-q}}{\left(d-1\right)^2}\left(\left\|\rho^{\Gamma}\right\|_1-1\right)^2,
\end{align}
where $\rho_i=|\psi_i\rangle\langle\psi_i|$. The first inequality is from (\ref{theorem4}) and (\ref{theor5}), the second inequality is obtained from the convexity of the function $\mathit{f}\left(x\right)=x^2$, the last inequality is due to the convex property of the trace norm and $\|\rho^\Gamma\| \geq 1$ in (\ref{text1}).

From (\ref{text1}), similar to (\ref{theorem4}), (\ref{theor5}) and (\ref{text9}), we have that
\begin{equation}\label{text10}
C_q\left(\rho\right)\geqslant\frac{1-d^{1-q}}{\left(d-1\right)^2}
\left(\left\|\mathcal{R}\left(\rho\right)\right\|_1-1\right)^2
\end{equation}
in the cases of $q\geqslant2$ with $d\geqslant3$ and $q\geqslant3$ with $d=2$. Combining (\ref{text9}) and (\ref{text10}), we can obtain (\ref{text101}). Similarly, from (\ref{case1}) one can obtain (\ref{corbound4}). $\hfill\qedsymbol$

Theorem 1 gives tight lower bounds of the entanglement measure $C_q\left(\rho\right)$.
In \cite{PhysRevA.103.052423} a lower bound of $C_q\left(\rho\right)$ has been derived,
\begin{equation}\label{prab}
C_q\left(\rho\right)\geqslant\frac{\left[\max\left\{\|\rho^{\Gamma}\|_1^{q-1},
\|\mathcal{R}\left(\rho\right)\|_1^{q-1}\right\}-1\right]^2}{d^{2q-2}-d^{q-1}}.
\end{equation}
In the next section we calculate the $q$-concurrence of isotropic states
and show that our lower bound is tighter than the one given in (\ref{prab}).

\section{$q$-concurrence for Isotropic states}\label{qisotropics}

The isotropic states $\rho_F$ are the convex mixtures of the maximally entangled state and the maximally mixed state \cite{PhysRevA.59.4206},
\begin{equation}
\rho_F=\frac{1-F}{d^2-1}\left(I-|\Psi^+\rangle\langle\Psi^+|\right)
+F|\Psi^+\rangle\langle\Psi^+|,
\end{equation}
where $\ket{\Psi^+}=\frac{1}{\sqrt{d}}\sum_{i=1}^d\ket{ii}$ is the maximally entangled pure state, $I$ is the identity operator and $F$ is the fidelity of $\rho_F$ with respect to $\ket{\Psi^+}$, $F=\braket{\Psi^+|\rho_F|\Psi^+}$, $0\leqslant F\leqslant 1$.
$\rho_F$ is separable for $F\leqslant 1/d$ and invariant under the operation $U\otimes U^*$ for any unitary transformation $U$ \cite{PhysRevA.59.4206}. The entanglement of formation \cite{PhysRevLett.85.2625}, tangle and concurrence \cite{PhysRevA.67.012307}, and R$\mathrm{\acute{e}}$nyi $\alpha$-entropy entanglement \cite{PhysRevA.93.022324} for the isotropic states have studied. Furthermore, it has proven that $\|\rho_F^{\Gamma}\|_1=\|\mathcal{R}\left(\rho_F\right)\|_1=dF$ for $F>1/d$ \cite{rudolph2005further,PhysRevA.65.032314}.

By straightforward calculation the $q$-concurrence for $\rho_F$ is given by
\begin{equation}
C_q\left(\rho_F\right)=co\left(\xi\left(F,q,d\right)\right),
\end{equation}
where $co\left(\xi\left(F,q,d\right)\right)$ denotes the largest convex function that is upper bounded by the function $\xi\left(F,q,d\right)$, and
\begin{equation}\label{example10}
\xi\left(F,q,d\right)=1-\gamma^{2q}-\left(d-1\right)\delta^{2q},
\end{equation}
where $\gamma=\sqrt{F}/\sqrt{d}+\sqrt{\left(d-1\right)\left(1-F\right)}/\sqrt{d}$, $\delta=\sqrt{F}/\sqrt{d}-\sqrt{1-F}/\sqrt{d\left(d-1\right)}$ \cite{PhysRevA.103.052423}.

To show the tightness of our lower bound (\ref{text101}), let us first consider the case of $q=3$.

(i) $d=2$. (\ref{example10}) becomes
\begin{align}\label{example2}
\xi\left(F,3,2\right)=\frac{3}{4}\left(2F-1\right)^2,
\end{align}
where $F\in(1/2,1]$. As the second derivative of $\xi\left(F,3,2\right)$ with respect to $F$ is non-negative, we have
\begin{equation}\label{example3}
C_3\left(\rho_F\right)=\begin{cases}
0,       & F\leqslant 1/2,\\[2mm]
\displaystyle\frac{3}{4}\left(2F-1\right)^2,  &1/2< F\leqslant 1,
\end{cases}
\end{equation}
which is just our lower bound (\ref{text101}).
Therefore, for $q=3$ and $d=2$ our lower bound (\ref{text101}) is just the exact value of the $q$-concurrence for any two-qubit isotropic state $\rho_F$.
While (\ref{prab}) gives rise to
\begin{equation}\label{q3exc3pra}
C_3\left(\rho_F\right)\geqslant\frac{\left(2F+1\right)^2}{12}\left(2F-1\right)^2,
\end{equation}
whose lower bound is less than the exact value of $q$-concurrence.

In fact, from (\ref{bound1}) we have $f\left(2\right)=f\left(3\right)$ for $d=2$. Hence, $C_2\left(\rho_F\right)=2/3C_3\left(\rho_F\right)$, which is consistent with the 2-concurrence of any two-qubit isotropic state $\rho_F$ \cite{PhysRevA.67.012307}. This implies that our lower bound (\ref{text101}) is exact for both $C_2\left(\rho_F\right)$ and $C_3\left(\rho_F\right)$.

\begin{figure}[t]
  \centering
  \includegraphics[width=8cm]{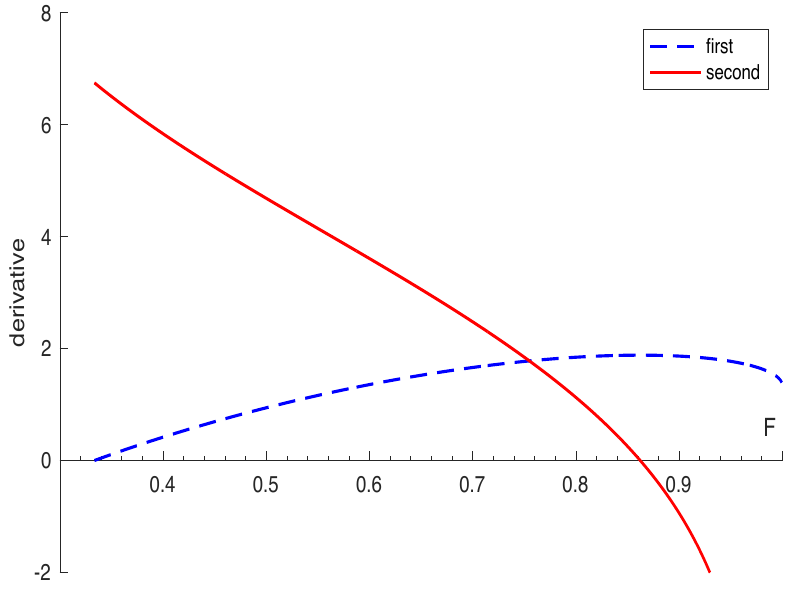}\\
  \caption{The first (dashed blue line) and the second (solid red line) derivatives of $\xi\left(F,3,3\right)$ with respect to $F$.}
  \label{f4}
\end{figure}

(ii) $d=3$. From (\ref{example10}) we have
\begin{equation}\label{text031}
\xi\left(F,3,3\right)=1-\gamma^6-2\delta^6
\end{equation}
for any $F\in(1/3,1]$, where $\gamma=\sqrt{F}/\sqrt{3}+\sqrt{2-2F}/\sqrt{3}$ and $\delta=\sqrt{F}/\sqrt{3}-\sqrt{1-F}/\sqrt{6}$. As the first derivative of $\xi\left(F,3,3\right)$ with respect to $F$ is always positive, $\xi\left(F,3,3\right)$ is monotonically increasing in the regime where $\rho_F$ is entangled, see Fig. \ref{f4}.
Since the second derivative of $\eta\left(F,3,3\right)$ with respect to $F$ becomes negative when $F\geqslant 0.86$, $\eta\left(F,3,3\right)$ is no longer a convex function for $F\in[0.86,1]$. As $C_3\left(\rho_F\right)$ is the largest convex function that is upper bounded by (\ref{text031}), we connect the point $F=0.86$ with the point $F=1$ by a straight line. Therefore, we obtain, see Fig. \ref{f4},
\begin{equation}\label{c33}
C_3\left(\rho_F\right)=\begin{cases}
0,       & F\leqslant 1/3,\\
\xi\left(F,3,3\right), &1/3< F\leqslant0.86,\\
1.777F-0.888,  &0.86< F\leqslant 1.
\end{cases}
\end{equation}

\begin{figure}[t]
   \centering
   \includegraphics[width=8cm]{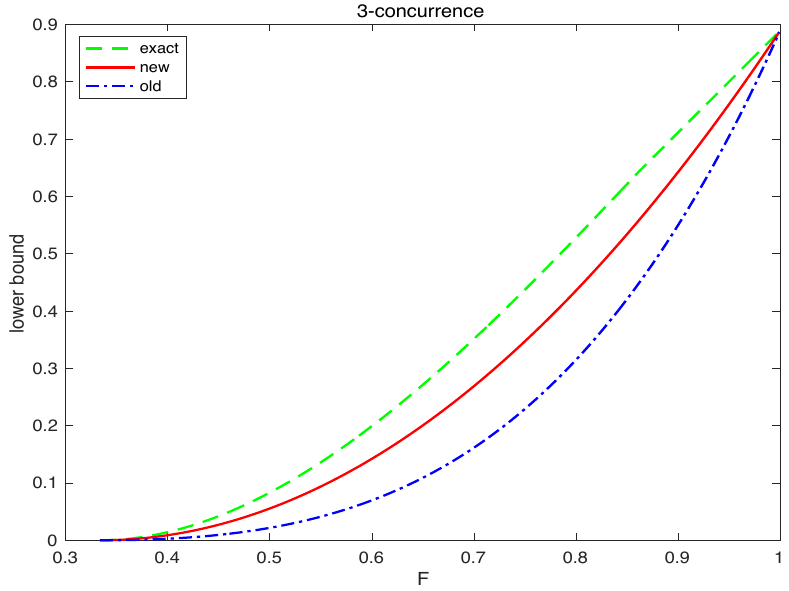}\\
   \caption{$C_3\left(\rho_F\right)$ versus $F$ for $d=3$. The dashed (green) line stands for $C_3\left(\rho_F\right)$ in (\ref{c33}). The solid (red)  line is for the lower bound (\ref{exc3}), the dotted dashed (blue) line is for the lower bound (\ref{exc3pra}).}
  \label{f5}
\end{figure}

From Theorem 2 we get that
\begin{equation}\label{exc3}
C_3\left(\rho_F\right)\geqslant\frac{2}{9}\left(3F-1\right)^2.
\end{equation}
While the lower bound (\ref{prab}) gives rise to
\begin{equation}\label{exc3pra}
C_3\left(\rho_F\right)\geqslant\frac{\left(3F+1\right)^2}{72}\left(3F-1\right)^2.
\end{equation}
Our lower bound of (\ref{exc3}) is tighter than (\ref{exc3pra}), see Fig. \ref{f5}.

For the case $q=4$, we have when $d=2$,
\begin{align}\label{example4}
\xi\left(F,4,2\right)=\frac{8-\left(2F-1\right)^2}{8}\left(2F-1\right)^2
\end{align}
for any $F\in(1/2,1]$. As the second derivative of $\xi\left(F,4,2\right)$ with respect to $F$ is non-negative, we obtain
\begin{equation}\label{example5}
C_4\left(\rho_F\right)=\begin{cases}
0,       & F\leqslant 1/2,\\[2mm]
\displaystyle\frac{8-\left(2F-1\right)^2}{8}\left(2F-1\right)^2,  &1/2< F\leqslant 1.
\end{cases}
\end{equation}

\begin{figure}[t]
  \centering
  \includegraphics[width=8cm]{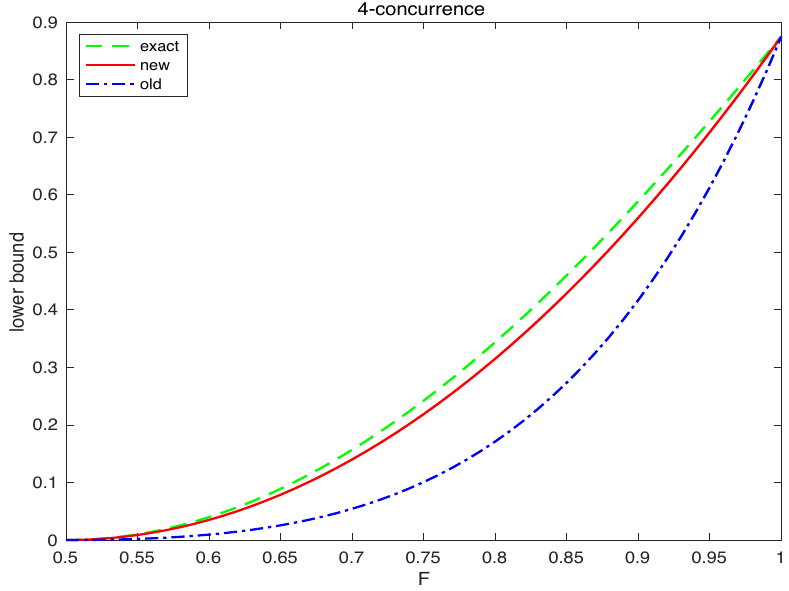}\\
  \caption{$C_4\left(\rho_F\right)$ versus $F$ for $d=2$. The dashed (green) line is for (\ref{example5}). The solid (red) line is for the lower bound of (\ref{example42}), and the dot dashed (blue) line is for the lower bound of (\ref{example42pra}).}
  \label{examf}
\end{figure}

From Theorem 2, we have
\begin{equation}\label{example42}
C_4\left(\rho_F\right)\geqslant\frac{7}{8}\left(2F-1\right)^2.
\end{equation}
While from (\ref{prab}), one gets
\begin{equation}\label{example42pra}
C_4\left(\rho_F\right)\geqslant\frac{\left(4F^2+2F+1\right)^2}{56}\left(2F-1\right)^2.
\end{equation}
Obviously, our lower bound (\ref{example42}) is tighter than (\ref{example42pra}),
see Fig. \ref{examf}.

\section{conclusion}\label{conclusion}
We have derived tighter lower bounds of the $q$-concurrence for $q\geq 2$ with $d\geq 3$ and $q\geqslant s$ with $d=2$. Moreover, we calculated the $q$-concurrence for isotropic states.
In particular, we obtained the analytical formulae of the $q$-concurrence for isotropic states with $q=3$ and $d=2,3$, as well as with $q=4$ and $d=2$. It turned out that our lower bound is exact for $q=2,3$ and $d=2$. These results may highlight further investigations on implications of quantum entanglement to quantum information processing.

\bigskip
\section*{Acknowledgements}
This work was supported by the National Natural Science Foundation of China (Nos. 12075159, 12171044); Beijing Natural Science Foundation (Grant No. Z190005); Academy for Multidisciplinary Studies, Capital Normal University; Shenzhen Institute for Quantum Science and Engineering, Southern University of Science and Technology (Grant No.SIQSE202001); the Academician Innovation Platform of Hainan Province.

\section*{Declarations}
All data generated or used during the study appear in the submitted article.


\end{document}